\documentclass[usenatbib]{mnras}

\usepackage[T1]{fontenc}
\usepackage{ae,aecompl}

\usepackage{graphicx}	
\usepackage{amsmath}	
\usepackage{amssymb}	
\usepackage{makecell} 
\usepackage{soul}
\defcitealias{KT18}{KT18}
\usepackage{bm}
\newcommand{\gradZ}{ \left\langle \nabla Z \right\rangle }
\newcommand{\Hii}{H\textsc{ii} }
\newcommand{\NSH}{N$_2$S$_2$H$\alpha$}

\newcommand{\ON}{O$_3$N$_2$}

\title[The ``bubbly" ISM model]{The ``bubbly" interstellar medium as origin for the inhomogeneous internal metallicity distributions in large disk galaxies}

\author[Metha et al.]{
Benjamin Metha$^{1,2}$\thanks{\hbox{benmetha@caltech.edu}}, Michele Trenti$^{1}$, Colin Norman$^{3,4}$
\\
$^1$School of Physics, The University of Melbourne, VIC 3010, Australia\\
$^2$IPAC, California Institute of Technology, 1200 E. California Blvd., Pasadena, CA 91125, USA\\
$^3$Center for Astrophysical Sciences, Department of Physics and Astronomy, The Johns Hopkins University, 3400 N Charles St. Baltimore, MD 21218, USA\\
$^4$Space Telescope Science Institute, 3700 San Martin Drive, Baltimore, MD 21218, USA\\
}

\date{Accepted XXX. Received YYY; in original form ZZZ}

\pubyear{2025}

\begin{document}
\label{firstpage}
\pagerange{\pageref{firstpage}--\pageref{lastpage}}
    \maketitle

\begin{abstract}
Resolved metallicity studies of local disk galaxies have revealed that their interstellar media (ISMs) are far from chemically homogeneous, displaying significant ($\sim 0.05$ dex) variations in the metallicity on characteristic scales of a few hundred parsecs. Such data is at odds with most analytical models, where the ISM is predicted to be more well-mixed. Here, we suggest that the observed small-scale features seen in galaxies may be superbubbles of metal-enriched gas created by a collection of core collapse supernovae with tight spatial (and temporal) correlation. 
In this scenario, the size of the metallicity fluctuations (superbubble radius, $\phi$) is set by the disk scale height of the galaxy in question (after which point shock breakout favours preferential expansion along directions perpendicular to the dense disc), and the amount of additional metals contained within a fluctuation is proportional to the star formation efficiency in superbubble regions ($\epsilon$). To test this theory, we analysed metallicity maps from the PHANGS-MUSE sample of galaxies using a geostatistical forward-modelling approach. We find $\phi \simeq 300$ pc and $\epsilon = 0.1-0.2$, in good agreement with our theoretical model. Further, these small-scale parameters are found to be related to the global galaxy properties, suggesting that the local structure of the interstellar medium of galaxies is not universal. Such a model of star formation paints a new picture of galaxy evolution in the modern universe: in large local galaxies, star formation appears steady and regular when averaged over large scales. However, on small scales, these large galaxies remain intrinsically bursty like their smaller, high-redshift counterparts.
\end{abstract}



\section{Introduction} \label{sec:intro}

The ways that galaxies form and evolve over time is well-understood on large scales as the result of an overdense region accreting gas. However, there are many open questions on this process on small scales. In particular, it is known that star formation must be regulated by stellar feedback, wherein stars and their products release energy into the interstellar medium (ISM) that delays the collapse of future generations of stars. However, virtually every aspect of this phenomenon is still uncertain, leaving theoretical and numerical models of galaxy formation on an unstable footing. Understanding the structure of the ISM, and the competing forces that affect it, is one of the major theoretical challenges in galaxy evolution \citep{Naab+17}.

Resolved distributions of metals (elements heavier than Helium, released by stars throughout ad at the end of their lifetimes) in galaxies can be used to shed light on the internal processes that shape the ISM of these galaxies \citep{Tremonti+04, Zahid+12, Zahid+14, Gibson+13, Ho+15, Belfiore+19, Franchetto+21}. Oxygen, the most commonly observed metal, is formed in stars and released into the interstellar medium through core-collapse supernovae \citep{Hashimoto+18}. Some fraction of this metal-enriched gas will mix back into the ISM of galaxies, while some hot gas will escape the galaxy disc to enter the circumgalactic medium before either falling back onto the galaxy as a galactic fountain, or escaping the galaxy as a metal-enriched outflow -- however, the fraction of gas in a galaxy that takes each of these routes, and what large scale properties of galaxies regulate this, is not well established 
\citep{Christensen+18}. Similarly, it is known that star forming galaxies must accrete cold gas to resupply the fuel for forming stars, but it is not clear whether this happens preferentially on the central regions or the outskirts of galactic discs \citep{Schonrich+McMillan17}. Furthermore, there is still ongoing debate about whether star formation happens only in the densest regions of galaxies,
 or over a range of densities owing to the fractal structure of the ISM \citep{Wright2020}.

With the modern generation of high resolution integral field units (IFUs), it is now possible to capture details about the internal metallicity distribution of galaxies with resolutions of tens of parsecs \citep[e.g.][]{Erroz-Ferrer+19, Lopez-Coba+20, Emsellem+22}. Such small-scale physical observations are crucial for shedding light on the physical processes that drive galaxy evolution, and how they relate to the larger scale properties of galaxies, which is a major focus of current astrophysical research \citep{NASA2020}.

Variations within the local metallicity distribution of galaxies have been observed for a long time. In \citet{Belley+Roy92}, using the [O III]/H$\beta$ metallicity diagnostic of \citet{Edmunds+Pagel84}, azimuthal metallicity variations by a factor of 2 were observed for two large face-on spiral galaxies, NGC 629 and NGC 6946. 
Examination of metallicity maps constructed with strong emission line based metallicity diagnostics from recent IFU surveys has confirmed these significant chemical inhomogeneities in galaxies \citep[e.g.][]{Sanchez-Menguiano+18, Ho+17, Vogt+17}. Even when direct methods are used to fit metallicity profiles for a galaxy, variations in the metallicity profiles of these galaxies are still observed (Section \ref{sec:evidence_of_inhomogeneities}).

Such metallicity variations have been challenging to explain with analytical theories. Assuming a galaxy has a constant star formation history that is uniform in space, \citet{Roy+Knuth95} argue that galaxies ought to be far more homogenous than those observed. Based on a timescales argument, \citet{Scalo+Elmegreen04} reach the same conclusion. By considering an inhomogeneous chemical evolution model based on overlapping regions of enriched material, \citet{Oey03} found that even in the limit of no mixing, the spread in stellar metallicities observed in our Galaxy is greater than the expected range of stellar metallicities. In more recent years, the statistical metallicity fluctuation model of \citet{KT18} was found to underestimate the variance of metallicity fluctuations caused by sub-kpc processes by a factor of $10^3 - 10^4$ when compared to data from the PHANGS sample \citep{Metha+21}. 


Several solutions to this disagreement between data and theory have been proposed. \citet{Roy+Knuth95} put forward two possible solutions to this problem -- that these metallicity fluctuations may be caused by supernova bubbles triggering additional star formation on at the sites of enrichment (before the chemical inhomogeneities have been mixed in to the ISM), or by the infall of clouds of pristine gas into a galaxy. 

In this work, based on high-resolution metallicity maps produced using state-of-the-art IFU observations, we propose a different explanation for these large chemical inhomogeneities seen in galaxies.
We suggest that these $\sim 200$ pc, $\sim0.05$ dex metallicity fluctuations are the face-on signatures of superbubbles,
caused by highly spatially and temporally correlated core-collapse supernovae -- the final fates of highly efficient star-formation events, as described in \citet{Norman+Ikeuchi89}. After reviewing the literature for evidence for and against a well-mixed ISM in Section \ref{sec:evidence_of_inhomogeneities}, we introduce this model formally and extract order-of-magnitude parameter predictions for the characteristic spatial extent and amount of metal enrichment seen in these metallicity fluctuations in Section \ref{sec:model}. In Section \ref{sec:real_data}, we compare these predictions to the parameters inferred from a geostatistical analysis of the metallicity maps of 19 local star-forming spiral galaxies, constructed using data from the PHANGS team \citep{Emsellem+22}. We highlight evidence for connections between the parameters that govern the small-scale physics of star formation and a galaxy's global parameters in Section \ref{ssec:local-global}. Further predictions are made, and experiments are proposed that would be able to test them, in Section \ref{sec:future_obs}. We discuss our results in Section \ref{sec:discussion}, and summarise our most important findings in Section \ref{sec:conclusions}.







Throughout this text, we adopt the values of solar metallicity computed in \citet{Asplund+21} -- this is 0.0139 (as a mass ratio of elements heavier than Helium to all mass in a galaxy, using the definition more commonly used by theorists), or $12+\log([$O/H$]) = 8.69$ (in units used by observers).

\section{Evidence for the existence of metallicity fluctuations}
\label{sec:evidence_of_inhomogeneities}

Whether or not the ISM of galaxies is well-mixed or inhomogeneous is an ill-defined question that is hotly debated \citep{Esteban+22}. Using the direct method to measure metallicities of \Hii regions, \citet{Bresolin11} found a $0.06$ dex scatter in the metallicity profile of M33. Similarly, the CHAOS (CHemical Abundance Of Spirals) have used the direct method to quantify the degree of homogeneity in the ISM of five nearby galaxies (M101, NGC 628, NGC 2403, NGC 3184, and NGC 5194), finding scatters in metallicity of 0.04 to 0.1 dex \citep{Croxall+16, Berg+20, Rogers+21}.

In all of these studies, it is not clear whether the scatter in the observed metallicity distribution are due to observational errors, or true (physical) metallicity variations about a mean radial trend. Using techniques from geostatistics, \citet{Metha+21} decomposed the residuals of the metallicity distribution of eight galaxies observed by the PHANGS collaboration into spatially correlated and uncorrelated components. Using three different strong emission line diagnostics, they found that spatially correlated variance accounts for approximately $50\%$ of the variance in the metallicity profiles of galaxies, indicating that these galaxies possess a spatially-varying component to their measured metallicity fields beyond their metallicity gradients. 

While these two results may appear to be at odds with each other, the tension is small. The fluctuations found in \citet{Metha+21} and in this work are on the order of $\sim 0.05$ dex, which is within the $0.04-0.1$ dex range of scatter quoted by direct method observations of nearby galaxies. This would imply that half to all of the variance in metallicity in galaxies observed with the direct method would be the consequence of real, spatially correlated metallicity inhomogeneities, and not measurement error.

With a single emission line diagnostic, it may be possible that some of the spatially-correlated variation in strong emission line diagnostics may be due to other spatially correlated nuisance parameter such as ionisation parameter varying over the field of view. To ensure that our results are robust against these nuisance variables, we use three different metallicity diagnostics in this work, each with their own strengths and weaknesses (Section \ref{sec:real_data}).

Throughout the rest of the paper, we assume that our strong emission line based metallicity diagnostics trace variations in the metallicity, and not another quantity. While direct methods show that galaxies are homogeneous to $\sim 0.1$ dex, we will show that our observed sample of galaxies display chemical inhomogeneities that are smaller than this value.

\section{The Model}
\label{sec:model}

\begin{figure*}
    \centering
    \includegraphics[width=0.8\linewidth]{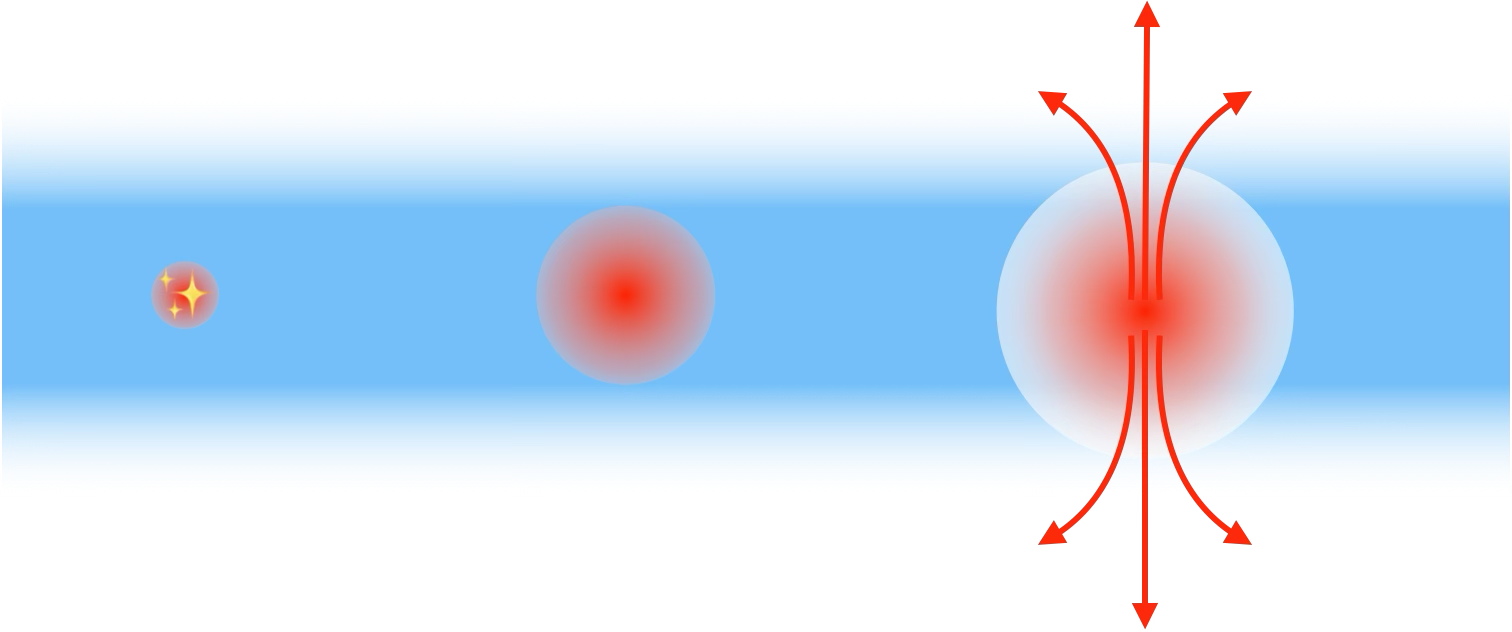}
    \caption{An illustration of the ISM model we propose as the explanation for the metallicity fluctuations observed in \citet{Metha+21}, inspired by the "chimney" model of \citet{Norman+Ikeuchi89}. \textit{Left:} A burst of star formation occurs in a dense pocket of gas, leading to $10-100$ core collapse supernovae occurring at a similar location and time. \textit{Centre:} These supernovae create a spherical region of expanding hot metal-enriched gas, shown in red. This sphere of hot gas expands until it reaches the scale height of the galaxy (\textit{right}), after which point the hot gas preferentially expands away from the high-density galaxy disc. This process would leaves a metal-enriched pocket of gas as an observable, with the lower metallicity ambient gas of the ISM blown away from the site of the supernovae.}
    \label{fig:cartoon}
\end{figure*}

To demonstrate the expected scaling relations and order-of-magnitude parameter values of our model, we consider a main sequence galaxy with a stellar mass of $10^{10.5} M_\odot$, a dark matter halo mass of $10^{12}M_\odot$, and a star formation rate of $1 M_\odot $ yr$^{-1}$. The stellar mass was chosen to be close to the median mass of the PHANGS sample of galaxies. The dark matter halo was computed using the halo-stellar mass relation of \citet{Behroozi+13}, and the star formation rate was chosen to agree with the star formation main sequence defined in \citet{Saintonge+Catinella22}.\footnote{Using Equation 7 of \citet{Saintonge+Catinella22}, our model galaxy is actually 0.1 dex below the main sequence, which is well within the $0.2-0.3$ dex observed scatter of the main sequence.} We assume a cold gas mass of $10^9 M_\odot$, consistent with a depletion time of $\tau_{\text{dep}} = 1$ Gyr, a typical value for spiral galaxies in the local Universe \citep{Tacconi+13, Saintonge+Catinella22}. Morphologically, we assume the gas lies in an azimuthally-symmetric exponential disc, with an effective radius of $10$ kpc and a scale height for the cold gas component of $200$ pc, chosen to match the scale height of the thin stellar discs found in \citet{Comeron+18}, assuming the scale height of the cold gas component of a galaxy matches the scale height of the dynamically cold thin disc. 

Within this medium, a positive metallicity fluctuation is born from a cluster of high-mass stars (10-100 stars, comparable to the number of stars in an OB association; \citealt{Norman+Ikeuchi89}) all exploding as supernova within a characteristic massive star evolutionary timescale of $\sim 10^6$ years within a star forming region of a$\sim 10$ parsec in size (see Fig~\ref{fig:cartoon} - left panel). The supernovae create a superbubble with an energy of $10^{54}$ ergs (Fig~\ref{fig:cartoon} - middle panel), sufficient to eventually break out of the galaxy (Fig~\ref{fig:cartoon} - right panel). By integrating a \citet{Kroupa_IMF} IMF between $8 M_\odot$ (the smallest mass a supernova can be; \citealt{Smartt09}) and $150 M_\odot$ (the mass of the largest stars that can form in the local Universe), we get that $0.69\%$ of stars end their lives in a core collapse supernova. 
Therefore, a star formation event where $1500 - 15000 M_\odot$ of material is converted into stars is required to generate a superbubble. Considering that such superbubbles are formed at a rate of $\gamma_{\text{SB}} \sim 7 \times 10^{-5}$ yr$^{-1}$ \citep{Norman+Ikeuchi89}, this would imply that star formation rates of $0.1-1 M_\odot$ yr$^{-1}$ are occurring in molecular clouds in which superbubbles will be born -- that is, $10-100\%$ of the star formation of a galaxy is highly spatially clustered. This is observationally supported to order of magnitude, as a fraction of $\sim 30\%$ of high mass stars formed in a galaxy are associated with \Hii regions \citep{Zhang+20}.

Such superbubbles will expand until they reach approximately 
the scale height of the galaxy, whereupon they break out, releasing energy and enriched gas into the circumgalactic medium. Viewed face-on, such superbubbles would appear as enriched regions of interstellar gas, with radii similar to the scale height of the galaxy. 

The amount of metals released into the interstellar medium depends on the efficiency, $\epsilon$, at which gas is converted into stars as well as on the yields of the supernovae. We take the IMF-weighted yield of core-collapse supernovae to be $y=0.015$, the value computed by \citet{Kobayashi+20} for gas of approximately solar metallicity ($Z=0.02$). Then, the amount of new metals released into the ISM by one superbubble is:

\begin{equation}
    M_{\text{metals}} = \epsilon \times y \times M_{\text{gas}}
    \label{eq:yield}
\end{equation}
where $M_{\text{gas}}$ is the mass of gas involved in the star-formation event that produced the core collapse supernova progenitors. From this equation, provided that the other gas within the area of a fluctuation is blown out by the superbubble's shock wave, the increase in metallicity in a superbubble will be equal to $M_{\text{metals}} / M_{\text{gas}}\approx \epsilon  y$. 
Using data from \citet{Behroozi+13b}, for a galaxy of stellar mass $10^{12} M_\odot$ at a redshift $z\sim 0$, a typical star formation efficiency (defined as the ratio between the star formation rate and the infall rate of intergalactic gas) is about $0.1$. With this value, we predict the amplitude of one of these metallicity fluctuations to be $\Delta Z = 0.0015$, or approximately $0.1Z_\odot$, corresponding to an observable fluctuation in the Oxygen abundance of $0.04$ dex.

The number of superbubbles present within a galaxy can be predicted based on superbubble formation rates and lifetimes. From \citet{Norman+Ikeuchi89}, the lifetime of a superbubble is $10^7$ yr. Combining this with a superbubble rate of $7 \times 10^{-5}$ yr$^{-1}$, we get a predicted number of superbubbles of $\sim 700$ per galaxy. This is in agreement with the number of star forming sites that are present within the discs of large, gas-rich galaxies ($\sim 10^3$, \citealt{Roy+Knuth95}). Assuming each superbubble occupies a region of radius equal to the scale height of the cold gas component of the galaxy ($\sim 200$ pc), one fluctuation takes up $4 \times 10^{-4}$ of the area of the galaxy's $10$ kpc disc. Combining these numbers, we predict that $\sim 28\%$ of the area of a galaxy would be covered by these superbubbles. This is consistent with the $\sim 10\%$ \Hii coverage fraction reported by \citet{Roy+Knuth95}.

\section{Comparisons to observations}
\label{sec:real_data}

To test the predictions of this model, we examined the 19 galaxies observed with VLT/MUSE as part of the Physics at High ANGular Resolution Survey (PHANGS), presented in \citet{Emsellem+22}. Properties of these galaxies are listed in Table \ref{tab:phangs}. All of these galaxies have been observed with a spatial resolution finer than 20 pc per pixel, far below the threshold criterion of $\sim 100$ pc per pixel required to do a geostatistical analysis on this data \citep{Metha+24}. 

\begin{table*}
    \begin{tabular}{lrrrrrrrr}
    Name & Distance (a) & Log(M*) (b) & Log(SFR) (b) & $\Delta$SFMS (b) & R25 (c) & PA (d) & i (d) & Pixel size\\
& [Mpc] & [$M_\odot$] & [$M_\odot$ yr$^{-1}$] & [dex] & [kpc] & [deg] & [deg] & [pc/pixel]\\
\hline
IC5332 & 9.0 & 9.67 & -0.39 & 0.01    & 7.9 & 74.4 & 26.9 & 8.7\\
NGC0628 & 9.8 & 10.34 & 0.24 & 0.18   & 14.0 & 20.7 & 8.9 & 9.5\\
NGC1087 & 15.9 & 9.93 & 0.12 & 0.33   & 6.9 & 359.1 & 42.9 & 15.4\\
NGC1300 & 19.0 & 10.62 & 0.07 & -0.18 & 16.6 & 278.0 & 31.8 & 18.4\\
NGC1365 & 19.6 & 10.99 & 1.23 & 0.72  & 34.2 & 201.1 & 55.4 & 19.0\\
NGC1385 & 17.2 & 9.98 & 0.32 & 0.50   & 8.5 & 181.3 & 44.0 & 16.7\\
NGC1433 & 18.6 & 10.87 & 0.05 & -0.36 & 16.8 & 199.7 & 28.6 & 18.1\\
NGC1512 & 18.8 & 10.71 & 0.11 & -0.21 & 23.0 & 261.9 & 42.5 & 18.3\\
NGC1566 & 17.7 & 10.78 & 0.66 & 0.29  & 18.5& 214.7 & 29.5 & 17.2\\
NGC1672 & 19.4 & 10.73 & 0.88 & 0.56  & 15.7 & 134.3 & 42.6 & 18.8\\
NGC2835 & 12.2 & 10.00 & 0.09 & 0.26  & 11.4 & 1.0 & 41.3 & 11.8\\
NGC3351 & 10.0 & 10.36 & 0.12 & 0.05  & 10.4 & 193.2 & 45.1 & 9.7\\
NGC3627 & 11.3 & 10.83 & 0.58 & 0.19  & 16.8 & 173.1 & 57.3 & 11.0\\
NGC4254 & 13.1 & 10.42 & 0.49 & 0.37  & 9.5 & 68.1 & 34.4 & 12.7\\
NGC4303 & 17.0 & 10.52 & 0.73 & 0.54  & 16.8 & 312.4 & 23.5 & 16.5\\
NGC4321 & 15.2 & 10.75 & 0.55 & 0.21  & 13.3 & 156.2 & 38.5 & 14.7\\
NGC4535 & 15.8 & 10.53 & 0.33 & 0.14  & 18.8 & 179.7 & 44.7 & 15.3\\
NGC5068 & 5.2 & 9.40 & -0.56 & 0.02   & 5.6 & 342.4 & 35.7 & 5.0\\
NGC7496 & 18.7 & 10.00 & 0.35 & 0.53  & 9.3 & 193.7 & 35.9 & 18.2\\ 
\end{tabular}
\caption{Table of PHANGS muse galaxy properties, taken from \citet{Emsellem+22}. References: (a): \citet{Anand+21}; (b): \citet{Leroy+21}; (c): from LEDA  \citep{HyperLEDA3}, converted to a physical size using the provided distance; (d): \citet{Lang+20}.}
\label{tab:phangs}
\end{table*}

Reduced maps with the intensity of a collection of emission lines from $4860-6740$\AA~ using the native resolution PSFs were downloaded from the PHANGS webpage.\footnote{\url{http://www.phangs.org/data}} A signal-to-noise cut of $10$ was applied to each line map. Extinction corrections from dust attenuation were applied by fixing the Balmer decrement to be H$\alpha/$H$\beta=2.86$ (asssuming Case B recombination) and using the extinction curve of \citet{ccm89}, assuming $R_V=3.1$ to match the Milky Way value. Spaxels were selected as belonging to star-forming (H\textsc{ii}) regions using both the Nitrogen-based and Sulphur-based BPT diagnostics of \citet{Kauffmann+03} and \citet{Kewley+01}. Only spaxels that were classified as belonging to star-forming regions by both of these diagnostics were considered for further analysis.

\begin{figure*}
    \centering
    \includegraphics[width=0.7\linewidth]{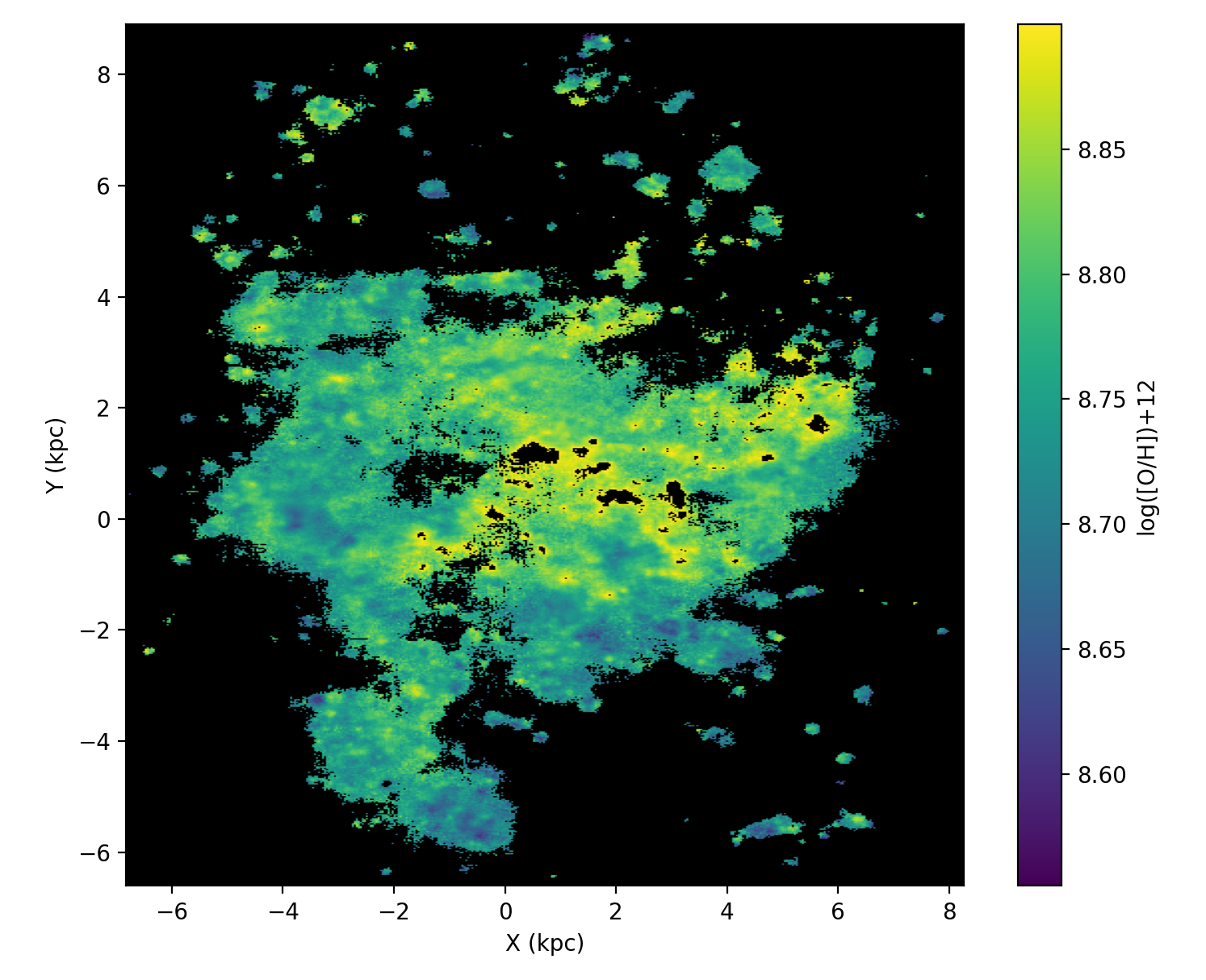}
    \caption{The metallicity distribution of NGC 1385, computed using the \ON\ metallicity diagnostic of \citet{Curti+17}. In this Figure, both a large scale trend in decreasing metallicity (corresponding to a negative metallicity gradient) and small-scale (sub-kpc) fluctuations (which we argue correspond to face-on superbubbles) can be seen.}
    \label{fig:example}
\end{figure*}

For each H\textsc{ii} spaxel, the metallicity of the gas was calculated using three different strong emission line diagnostics. We adopt the \ON\ diagnostic with the calibration of \citet{Curti+17} as our fiducial metallicity diagnostic (calibrated by comparing the emission line ratio to the metallicity of stacked SDSS spectra computed using the direct ($T_e$-based) method), and show results computed using the \NSH\ diagnostic with the calibration of \citet{Dopita+16} and the Scal diagnostic of \citet{Pilyugin+Grebel16} in Appendix \ref{ap:other_Z_diags}. We compute the uncertainty of each metallicity estimate from the uncertainty in each line flux using linear error propagation, following Equation B1 of \citet{Metha+22}. We show the metallicity distribution for NGC 1385, one of the PHANGS galaxies, in Figure \ref{fig:example}.

Using this methodology, the number of \Hii spaxels with metallicity measurements available for each of the PHANGS galaxies ranged from $6740$ to $154115$. In \citet{Metha+24}, it was shown that using a subsample of $500$ randomly-selected spaxels does not affect the recovered parameters of the Gaussian process governing the 2D spatial structure of chemical fluctuations in galaxies. Therefore, to speed up computations, subsamples of $500$ \Hii spaxels were utilized to estimate the length scales and amplitudes of metallicity fluctuations within each galaxy.

A four-parameter model was used to describe the true metallicity distribution of each galaxy, following \citet{Metha+22} and \citet{Metha+24}. The metallicity of a galaxy at a location $\vec{x}$ is described by a random field:

\begin{equation}
\label{eq:breakdown_true_metallicity}
Z(\vec x) = \mu(\vec x) + \eta(\vec x),
\end{equation}
where $\mu(\vec{x})$ is the expected (mean) metallicity at each location within the galaxy. This is modelled as being linearly dependant on the distance from a galaxy's centre, $r(\vec{x})$:
\begin{equation}
    \mu(\vec{x}) = Z_{\text{char}} +  \gradZ \cdot \left( r(\vec{x}) - r_{\text{char}} \right).
    \label{eq:z_grad}
\end{equation}
Here, $\gradZ$ is the metallicity gradient of the galaxy, and $Z_{\text{char}}$ is the metallicity at the \textit{characteristic radius} of $r_{\text{char}} = 0.4 R_{25}$ \citep{Zaritsky+94}. Choosing to define a metallicity gradient in this way helps to reduce the correlation between $\gradZ$ and the overall metallicity of the galaxy \citep{Metha+24}.

On top of this metallicity gradient, there exists small-scale variations in the metallicities of galaxies, denoted by $\eta(\vec x)$. This stochastic component has zero mean, and a stationary and isotropic covariance kernel of:
\begin{equation}
    \text{Cov}(\eta(\vec{x} + \vec{h}), \eta(\vec{x})) =\sigma^2  \exp \left( - \frac{h}{\phi}\right),
    \label{eq:matern_1/3}
\end{equation}
where $h= |\vec{h}|$. This kernel is parametrised by two parameters: $\sigma^2$, the amount of variability in metallicity that is caused by spatially correlated effects (such as superbubbles that have not yet mixed into the ISM -- not to be confused with observational uncertainties on the metallicity, which are not spatially correlated), and $\phi$, the spatial scale of correlated metallicity fluctuations. It is these two parameters that we are most interested in fitting for our model: $\sigma^2$ determines the amount of metal enrichment in a fluctuation (predicted to be $\Delta Z \approx 0.1 Z_\odot$), and $\phi$ is predicted to be approximately the scale height of a galaxy (predicted to be $\phi \approx 200$ pc).

To estimate these four parameters for each galaxy, the Python package \textsc{emcee} \citep{emcee} was used. To get prior predictions of $Z_{\text{char}}$ and $\gradZ$, these two parameters were fit for each galaxy using a weighted least-squares (WLS) approach, accounting for the observational errors in metallicity for each spaxel but ignoring any spatially-correlated variations. Normal priors were then imposed for $Z_{\text{char}}$ and $\gradZ$, with means equal to the values found with the WLS method, and standard deviations of $0.1$ dex for $Z_{\text{char}}$ and $0.02$ dex kpc$^{-1}$ for $\gradZ$. For $\sigma^2$ and $\phi$, gamma priors were used, as this distribution has support over only the positive numbers and prevents unphysical negative values from being drawn. Parameters were chosen for the gamma distribution of $\phi$ so that there was a $1\%$ a-priori probability of $\phi < 20$ pc (comparable to the size of a single pixel) or $\phi > 2$ kpc (at which scale azimuthal variations in metallicity are rarely seen in face-on local spiral galaxies). For $\sigma^2$, parameters of the gamma distribution were chosen such that there was a $1\%$ chance of $\sigma^2$ being ten times greater than or ten times smaller than the observed variance of $Z(\vec{x})$ (accounting for both observational uncertainty and intrinsic variability) for each galaxy.

Samples of the posterior distributions for these for parameters were computed by running \textsc{emcee} for $480$ steps with $72$ walkers. For each chain, the first $80$ samples were discarded as part of the burn-in run. Chains were examined for non-convergence. For six galaxies (NGC1087, NGC1512, NGC1566, NGC1672, NGC4303 and NGC7496), a longer burn-in period of $120$ steps was deemed necessary when the Scal metallicity diagnostic was used; therefore an additional $40$ samples were taken for these galaxies when analysing maps computed with this diagnostic. 


\subsection{Small-scale parameters for PHANGS galaxies}

The size of metallicity fluctuations $\sigma$ (units: dex) fitted for each galaxy was converted into intrinsic metallicity variations $\Delta Z$ (unitless) by computing the median galactocentric radius of data points in each sample, and using the computed values of $Z_{\text{char}}$ and $\gradZ$ to convert this radius into a median overall metallicity for the data. These were then divided by $y=0.015$ to compute the star-formation efficiency $\epsilon$ of areas where superbubbles are generated for each galaxy. 

\begin{figure*}
    \centering
    \includegraphics[width=\linewidth]{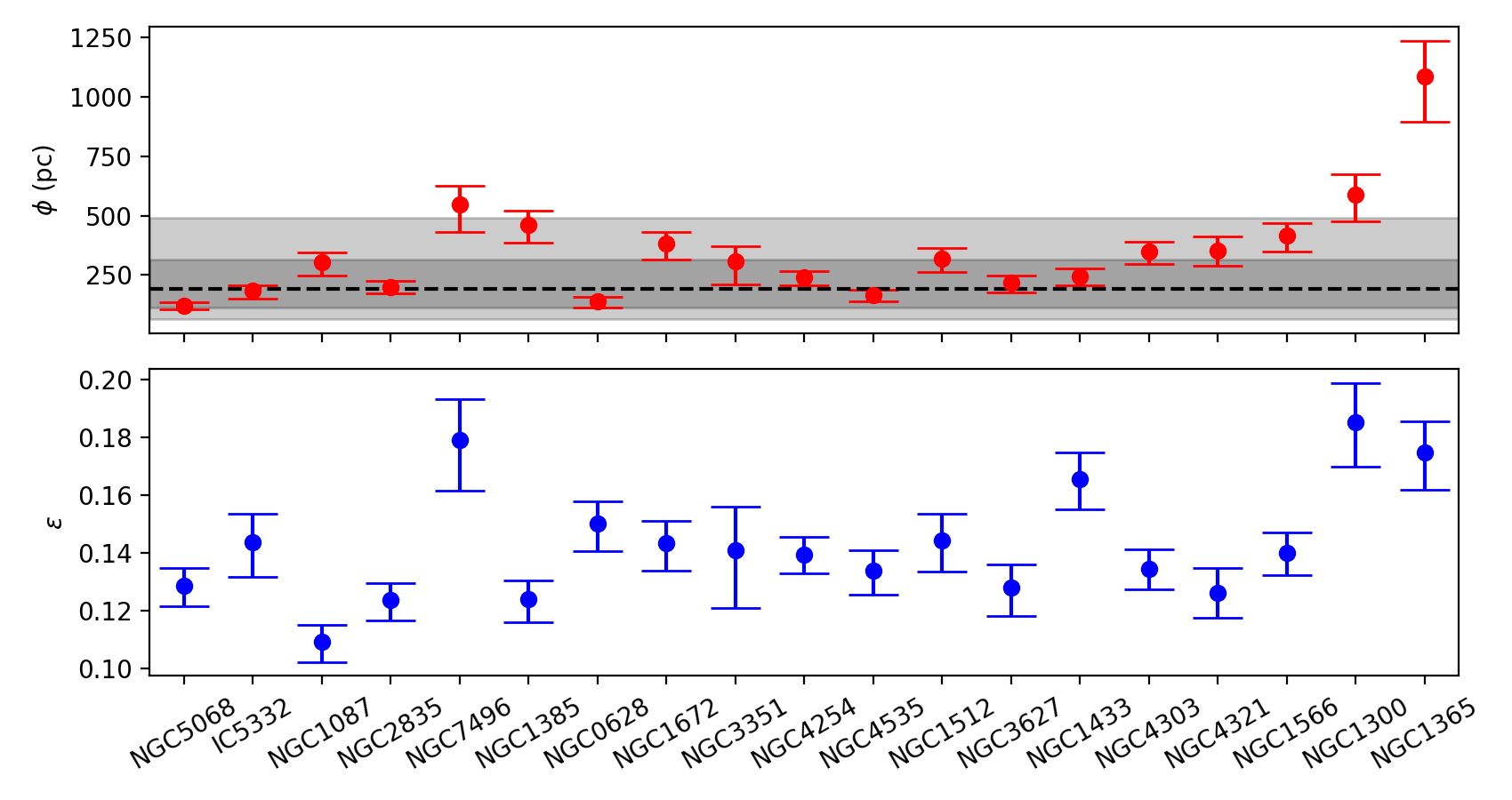}
    \caption{Small scale parameters for the metallicity fluctuations identified in the PHANGS galaxy sample, using the \ON\ metallicity diagnostic. The top panel shows how the characteristic size of fluctuations in a galaxy, $\phi$, changes throughout the galaxy sample. Galaxies here are ordered from the smallest (left) to greatest (right) stellar mass.
    The dashed black line in the top panel is the median scale height of thin stellar discs in the galaxy sample presented in Appendix B of \citet{Comeron+18}, with the dark- and light-shaded grey regions indicating the $1\sigma$ and $2\sigma$ spreads in this population. We see that values of $\phi$ are generally consistent with typical values of the scale heights of galaxies.
    In the lower panel, we plot the estimates of $\epsilon$, the star formation efficiency within the regions which become superbubbles. With this metallicity diagnostic, we find typical star formation efficiencies of $0.1-0.2$, indicating that highly efficient star formation occurred in order to produce the metallicity fluctuations observed.}
    \label{fig:smallscale_params}
\end{figure*}

We plot the spatial scale $\phi$ and $\epsilon$ computed for each of our galaxies, arranged by their stellar mass, in Figure \ref{fig:smallscale_params}. In the top panel, $\phi$ is plotted along with the median, $16-84$th percentile range, and $2.5-97.5$th percentile range of the stellar thin disc scale heights observed for a sample of 124 edge-on galaxies for which thin- and thick-stellar discs could be separately identified, presented in Appendix B of \citet{Comeron+18}. We find that for 11 of our 19 galaxies ($58\%$), the best-fitting value of $\phi$ falls within the $68\%$ confidence interval for scale heights expected from \citet{Comeron+18}, and 16 out of 19 ($84\%$) fall in the $95\%$ confidence interval. Overall, the median value of $\phi$ for the PHANGS sample was $308$ pc, a factor of 1.6 higher than the median scale height for the thin stellar disc of $191$ pc. We find two galaxies with significantly larger values of $\phi$ -- namely, NGC 1300 ($\phi = 587^{+111}_{-87}$pc) and NGC 1365 ($\phi = 1088^{+193}_{-149}$pc). For interpretation of our results, we note that these are the two most massive galaxies in our sample -- therefore, it is not unexpected that the scale heights would be larger in these galaxies.

For all galaxies in our sample, we see that the size of metallicity fluctuations is $\sigma=0.04-0.06$ dex, consistent with the limits of $0.04-0.1$ dex fluctuations seen in direct metallicity studies. Under our model, these fluctuations correspond to a value of $\epsilon= 0.1-0.2$, implying that $10-20\%$ of the gas mass within any given GMC region is being transformed into stars, the most massive of which explode as core collapse supernovae to power the expansion of the superbubble. This star formation efficiency is high, and comparable to the star forming efficiencies expected for large galaxies at high redshifts \citep{Shen+25}. From this, we  infer that the star formation of these larger galaxies is highly bursty on small scales, but such burstiness is averaged over several orders of magnitude when considering the entire galaxy's star formation history with a temporal resolution of $\sim 10^7$ years.

\subsection{Connection to global galaxy properties}
\label{ssec:local-global}

\begin{figure}
    \centering
    \includegraphics[width=\linewidth]{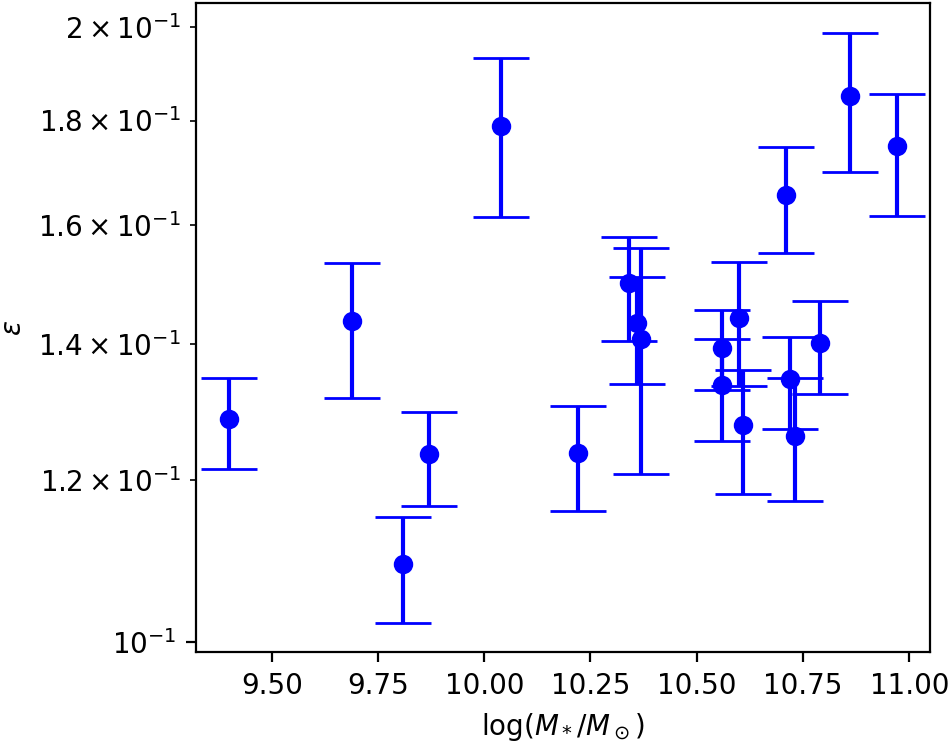}
    \includegraphics[width=\linewidth]{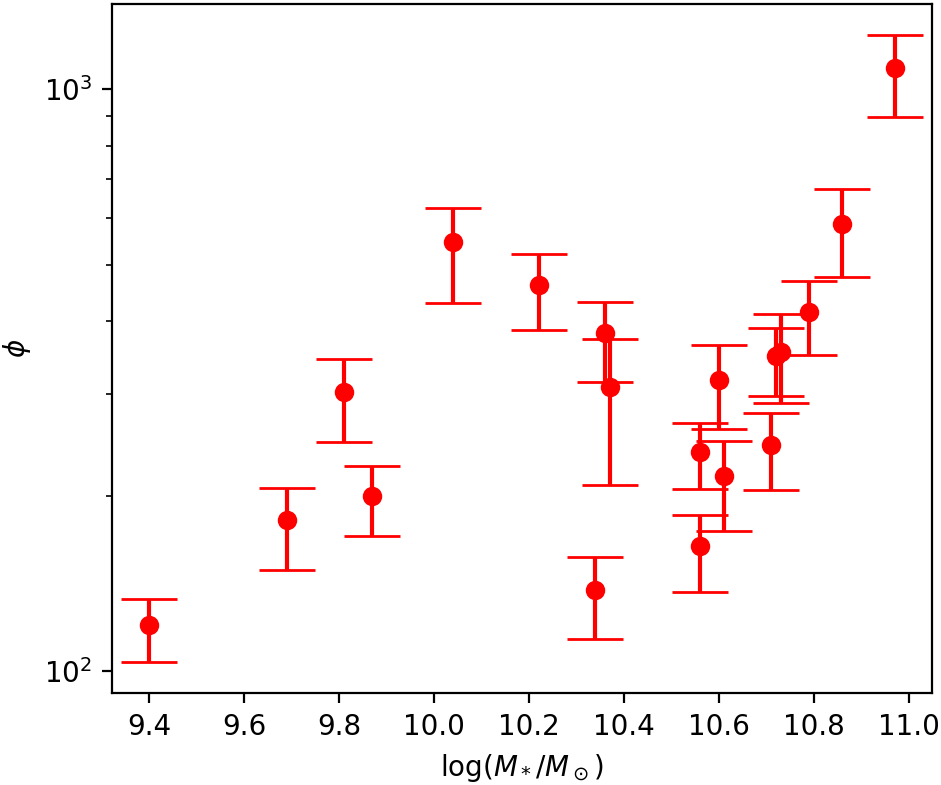}
    \includegraphics[width=\linewidth]{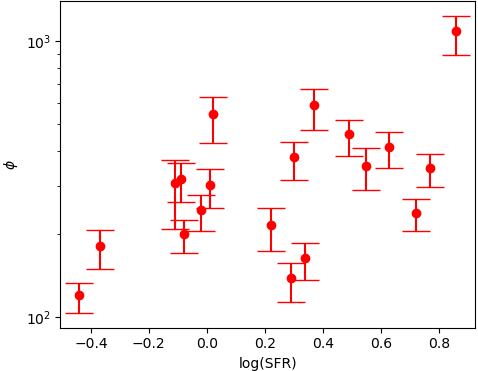}
    \caption{Moderate correlations can be seen between the small-scale properties of the metallicity fluctuations seen in galaxies and the global properties of the galaxies. In the top panel, we show the positive correlation between stellar mass and the star forming efficiency of gas in superbubbles, $\epsilon$ $(\rho=0.42)$. In the middle and lower panels, we show the positive correlations of $\phi$, the characteristic spatial scale of metallicity fluctuations, compared to two global galaxy properties: the stellar mass of galaxies (\textit{middle panel}; $\rho=0.53$), and their star formation rates (\textit{lower panel}; $\rho=0.54$).}
    \label{fig:local_and_global}
\end{figure}

To investigate the tentative relationship between $\epsilon$ and $\phi$, we explored whether both or either of these quantities depended on an additional parameter associated with our sample galaxies. We searched for sizeable 
positive or negative correlations (quantified by the Pearson correlation coefficient $\rho$) between the values of $\epsilon$ and $\phi$ recovered for these galaxies and the galaxies' overall (i) stellar mass, (ii) star formation rate, (iii) specific star formation rate, and (iv) distance above or below the main sequence. We found that the strongest of these global galaxy variables to correlate with $\epsilon$ was the stellar mass of the galaxies, with $\rho(\log(\epsilon), \log(M_*))=0.42$ for O3N2, and with similar values for the alternative metallicity diagnostics ($\rho = 0.4156$ for \NSH, and $\rho=0.3055$ for Scal). 
For $\phi$, we found significant correlations with two variables: the stellar mass of galaxies ($\rho(\log(\phi), \log(M_*))=0.53$ for \ON) and their star formation rates ($\rho(\log(\phi), \log$(SFR)$)=0.54$ for \ON). For all other pairs of small-scale and galaxy-scale parameters, no significant connection was seen between the variables consistently with all three metallicity diagnostics explored. 
We plot the key relationships between local and global parameters in Figure \ref{fig:local_and_global}.

The fact that the parameters regulating the strength and sizes of small-scale metallicity fluctuations are seen to depend on a galaxy's large-scale properties indicates that, to some degree, the physics of the interstellar medium is not universal between all galaxies, but rather that it is set by the properties of the galaxies as a whole. More specifically, each of these interdependencies between small-scale and global parameters can be explained physically.

We suggest that the relationship between $\epsilon$ and the stellar mass of galaxies could be a consequence of stellar feedback. For galaxies with stellar masses below $\log(M_*/M_\odot) \simeq 10.4$, the stellar mass to halo mass relation decreases with decreasing stellar mass, dropping by a factor of 2 between $\log(M_*/M_\odot) = 10.4$ and $\log(M_*/M_\odot) = 9.7$ at $z=0.1$ \citep{Behroozi+13}. This is due to star formation being less efficient for these lower mass galaxies as feedback from supernovae becomes an increasingly dominant term to quench star formation. Interestingly, we do not see a strong correlation between $\epsilon$ and the SFR of galaxies ($\rho = 0.127$), indicating that the local star formation efficiency of the ISM may be the same for galaxies with different star formation rates. In such a scenario, the differences in SFR between galaxies would be set by the number of starburst regions rather than by the efficiency of each one.

For the values of $\phi$ measured, the dependence on $M_*$ and SFR can be understood through the dependence of $\phi$ on the scale heights of galaxies. As galaxies grow to larger masses, their size increases, too. Assuming a constant or near constant flatness parameter for spiral galaxies, this growth translates to an increase in the scale heights of galaxies. The additional dependence on SFR may be due to galaxies with higher values of SFR having puffier discs due to the actions of star formation, or it may simply be due to a secondary dependence on mass through the star formation main sequence relation.

Overall, our results agree with those of \citet{Baker+23} -- that the structure of the ISM of a galaxy on small scales is, to some degree, determined by the galaxy's global properties.

\section{Predictions and future observables}
\label{sec:future_obs}

This model makes several predictions for the local structure of the ISM of galaxies, many of which are testable with existing galaxy data. In this Section, we outline several promising future directions for research that could either provide evidence for or against our predicted ISM model.


\textbf{$T_e$-based confirmation of metallicity diagnostics:} Perhaps the most important prediction of this model is that these metallicity fluctuations that we have analysed in this work really do exist. To confirm this, direct method metallicities would need to be collected for a similar sample of galaxies. This will be possible with new surveys, such as the SDSS-V Local Volume Mapper \citep{Drory+24}. Future 30m class telescopes will be able to take this even further, with their capability to resolve faint lines such as [N II]5755 or [S III]6312 with high spatial resolution. A geostatistical analysis will be able to search for spatial correlations between direct and strong emission line based metallicity measurements, to help better understand the relationship between metallicity, ionisation, and emission line ratios. 

\textbf{Spatial structure of star-formation}: There is a time lag between star formation events that create the stars that explode and deliver the metals, and the expansion of the super-bubbles to the sizes we see them at. Furthermore, energy injected by supernovae disrupts star formation, lowering the expected local star formation efficiency at the sites of the metal-rich bubbles. Therefore, we can't expect the metallicity to be locally correlated with the star formation rate. But we could expect the spatial structure of both the SFR and the metallicity to be similar. If the lifetime of a superbubble is $10^7$ years and H$\alpha$ surveys are sensitive to star formation over $10^7$ year timescales, then we would expect spatial correlation to be found between star formation rates observed with H$\alpha$ diagnostics and positive metallicity correlations. 

As an early investigation into whether the SFR profile of a galaxy exhibits spatial structure in a similar way to its metallicity map, we constructed two-point correlation functions for both the local SFR density and the metallicity (as computed with our fiducial \ON\ diagnostic) of our flagship galaxy, NGC 1385. 
Based on the commonly-used optical SFR tracer \citep[e.g.][]{Calzetti+07, Kennicutt+Evans12, Belfiore+23}, we model
$\log($SFR) as being linearly related to $\log($H$\alpha)$. Therefore, by fitting a radial linear profile to $\log($H$\alpha)$ and computing the residuals around this profile, we generate a map of the residuals around an exponential star formation profile ($\log($SFR) $ = a - b \cdot r$). We then construct a two-point correlation function from the resulting SFR residual map after filtering out all spatially uncorrelated noise to eliminate the effects of measurement error. A two-point correlation function of the metallicity residuals (with the metallicity gradient subtracted) was constructed in the same way.

\begin{figure}
    \centering
    \includegraphics[width=0.5\textwidth]{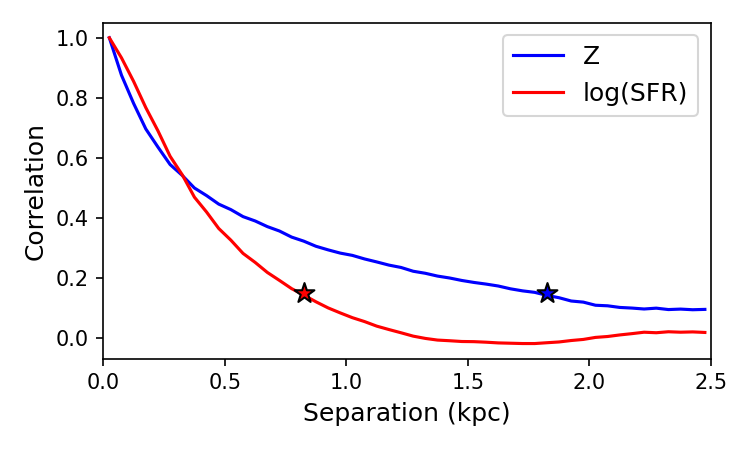}
    \caption{Two-point correlation functions revealing the spatial structure of the star formation profile residuals (red) and the metallicity residuals (blue) of NGC 1385, computed using H$\alpha$ brightness maps and the \ON\ diagnostic, respectively. Stars indicate the locations at which each of these correlation curves first drop below a value of $0.15$. We see that the correlation function for SFR approaches zero more rapidly than the metallicity correlation function, indicating that SFR is correlated on smaller spatial scales than metallicity, in line with our theory.}
    \label{fig:sfr_vs_z_2pc}
\end{figure}

We present and compare these correlation functions in Figure \ref{fig:sfr_vs_z_2pc}. From this plot, we see that the correlation function constructed for the star formation rate profile approaches zero faster than the correlation function for metallicity fluctuations. To quantify this observation, we compute the correlation lengths for each of these variables. To be consistent with the analysis of \citet{Zhang+25}, we define this length to be the distance at which a correlation curve first drops below a value of 0.15.\footnote{Note that this statistic is not $\phi$ as we defined it earlier.} We find that for the metallicity residuals, this occurs at a separation of $1.825$ kpc, whereas this value is only $825$ pc for the SFR maps. This indicates that star formation is more tightly spatially correlated than metallicity. Such a conclusion is consistent with our proposed model, as metals would be originating from tightly correlated regions of star formation before spreading out under the force of a superbubble explosion, followed by turbulence-driven diffusion.

We also computed the Pearson correlation coefficient between our residual log(SFR) and metallicity maps to be $\rho = 0.253$, suggesting that there is a spatial correlation between the resolved star formation rate density of galaxies and their local metallicity, contrary to the observations of the resolved fundamental metallicity relation by e.g. \citet{Baker+23}. However, this point cannot be made conclusively with analysis of one galaxy alone.
We leave a full investigation into the connection between the small-scale structures of metallicity and SFR involving a larger sample of galaxies to future work.

\textbf{Are holes seen in the dust?} If the measured metallicity fluctuations are the face-on signatures of expanding superbubbles, then our model predicts that these superbubbles must expel large amounts of cold gas to expand into the interstellar medium -- this is necessary to produce fluctuations in the metallicity that are on the order of what we see. If the cold gas were instead mixed in to superbubbles as they expand, then the metallicity fluctuations in the interstellar gas would be much smaller than what is observed. As a consequence of this, we predict that around the metal-enriched regions of galaxies, dust will have been blown out, forming shells that are the boundaries of these bubbles. Intriguingly,  holes that appear to be consistent with our model are seen in JWST infra-red images of the 19 PHANGS galaxies that are the subject of this work \citep{Lee+23}. We predict the spatial scales of these holes to be greater than, but comparable to the spatial scale of metallicity fluctuations; further we predict that the centres of the holes seen in the mid-IR dust maps to be coincident with the centres of regions with locally high gas-phase metallicities.

\textbf{Velocity signatures:} While a superbubble is expanding, it will leave a signature on the resolved velocity map of a galaxy, in which a small ($\sim 100$ pc) region of gas would appear to be moving towards the observer, with a large velocity width. Because both velocity and metallicity maps can be computed from the same IFU data, if such small-scale velocity fluctuations are observed, it would be possible to determine if their locations coincide with the locations of high metallicity regions. 

\textbf{Variations between arms and inter-arm regions:} In our model, superbubbles are born from a group of high-mass stars that are born at approximately the same time going supernova at approximately the same time. We identify OB associations as the progenitor systems for these superbubbles. Not only are these OB associations possibly visible in our sample of local galaxies \citep{Gouliermis+18}, but they are also known to be more concentrated along the spiral arms of a galaxy. This could explain the recent observations that galaxies are more metal-enriched along their spiral arms \citep{Bresolin+25}
To check for this signal, we could alternatively select samples of \Hii regions captured for on-arm and inter-arm regions, using the stellar arm maps of PHANGS galaxies computed by \citet{Querejeta+24}. We could then compare whether the parameters that govern the local metallicity structure of these galaxies are the same or different in these different regions. We would predict there to be more correlated metallicity fluctuations either on the spiral arms of galaxies where the density is the highest, or on the trailing edge of spiral arms, in the region that a density wave has most recently passed through, accounting for the time delay between the star formation event that efficiently creates an OB association, and the explosions of these stars $\sim 10^7$ years later.

\textbf{Changes in parameters between the inner and outer regions of galaxies}: The thicknesses of the gaseous discs of galaxies are expected to increase with galactocentric radius, a phenomenon known as \textit{disc flaring} \citep[e.g.][]{Padoan+01, Benitez-Llambay+18, Patra20}. If the size of metallicity fluctuations in a galaxy is tied to its scale height, it may be expected that $\phi$ should also increase with galactocentric radius. This could be tested by generating data sets by sampling over the inner/outer regions of the discs of galactic metallicity maps and comparing the recovered parameters in each subregion of the galaxies. We leave a proper analysis of the PHANGS data split into subregions in this way as a topic of future work.

\textbf{Can dwarf galaxies contain superbubbles?} It would be interesting to see if fluctuations like the ones predicted by our model for large galaxies were still present in dwarf galaxies; whether they have the self-gravity to retain the metals, or whether breaking through with ample momentum would significantly mix the gas, redistributing metals throughout the entire galaxy. If the latter is the case, this could be an explanation for the highly-bursty nature of star formation in early galaxies, where initial, strongly bursty star formation events have the power to completely disrupt a galaxy.


\section{Discussion} \label{sec:discussion}

In this study, we have used methods from geostatistics to extract information on the small-scale structure of metals in a sample of face-on galaxies. This methodology provides us with a richer set of statistics than can be extracted compared to current widely used methods such as the two-point correlation function \citep[e.g.][]{Li+21, Zhang+25}. While the latter only provides details on the correlation length scale of metallicity fluctuations, a geostatistical analysis also captures information about the amplitude of metallicity fluctuations. This additional information has allowed us to construct a theory that readily explains the origins of these inhomogeneities.

An analytical calculation of the expected amplitude of a metallicity fluctuation in a galaxy is given in \citet{Roy+Knuth95}. First, the fraction of a galactic disc that is enriched by a supernovae event compared to the total volume of a galaxy is computed to be $6 \times 10^{-6}$. Knowing that, and assuming a standard supernova rate of 1 per 100 years over a timescale of $10^{10}$ years, the total number of supernovae ($n$) that enrich any given point in the ISM is calculated to be $\sim 600$. The expected variation of metallicity, then, assuming only Poissonian statistical variation, is $Z_{\text{char}} / \sqrt{n}$. This leads to expected metallicity fluctuations that are of the order of $0.02$ dex -- approximately 2.5 times smaller than the fluctuations that we observe. 

To resolve this tension between data and theory, we relax the assumption that supernovae are spatially uncorrelated. If we instead consider supernovae to occur in spatially correlated clumps of star-forming gas, we find that superbubbles occupy a larger fraction of the disc ($\sim 10^{-4}$ for our galaxy model assumed in Section \ref{sec:model}), but occur less frequently, happening at a rate of $7 \times 10^{-5}$ yr$^{-1}$. Under this model, the number of enrichment events in each region of a galaxy is $n \sim 70$. 
With this smaller number of enrichment events, larger metallicity fluctuations of $0.05$ dex are expected -- in better agreement with the data.

In our model, star formation efficiencies of $\epsilon = 0.1-0.2$ were shown to be sufficient to explain the amount of metals present in the $\sim 0.05$ dex metallicity fluctuations seen in data. These efficiencies are in good agreement with the values given in \citet{Behroozi+13b}. At a redshift of zero, the highest star formation efficiency presented in this model was $\epsilon=0.22$ for galaxies with halo masses of $10^{11.5} M_\odot$. This halo mass is lower than the median halo mass for PHANGS galaxies of $10^{12.17} M_\odot$, computed by interpolating the stellar-mass halo-mass relation of \citet{Behroozi+13}. Based on the galaxies' stellar masses, the median galaxy-averaged star formation efficiencies of the PHANGS galaxies is predicted by \citet{Behroozi+13b} to be $0.06$ -- a factor of $\sim 3$ smaller than the star formation efficiencies derived for superbubble regions. We justify this discrepancy by arguing that superbubble progenitor regions are dense regions of the ISM where star formation efficiencies are higher. That is, on sub-kpc scales, we predict the ISM of large, local spiral galaxies to be bursty.

Such local burstiness may be hidden when galaxy data is spatially averaged. Considering that there are hundreds of positive metallicity fluctuations (corresponding to face-on superbubbles) in galaxies, a SED-based star formation history reconstructed for a galaxy will be the summation of hundreds of bursts, mimicking a more constant, less bursty signal. This is in line with the observation that larger galaxies in the local Universe have more constant star formation histories than dwarves, which are more bursty \citep[e.g.][]{Ting+25}.

It is natural to ask what fuels these local episodes of enhanced star formation. We suggest that one possible cause of enhanced star formation would be the infall of cold pockets of star-forming gas, as described by \citet{Roy+Knuth95}. Such an event would temporarily decrease the metallicity of a local region of the ISM, before forming a giant molecular cloud in which stars are born, including the O and B type stars that trigger the formation of a metal-enriched superbubble. Assuming efficient star formation, the timescale for this process is set by the lifetimes the giant molecular cloud, which is $\sim 10^7$ years, comparable to the lifetime of a superbubble \citep{Norman+Ikeuchi89}. This implies that a local region of the ISM undergoing bombardment from a cold gas clump would spend equal amounts of time with a lower local metallicity and an enhanced local metallicity, which naively makes sense considering the observation that there are comparable numbers of metal-rich and metal-poor subregions of galaxies. To create regions with $0.04$ dex lower metallicities than the surrounding ISM, clumps of gas of mass $\sim 10^5 M_\odot$ would need to be accreted.

If this process operates on a similar spatial scale to the size of observed superbubbles, then signatures of this phenomenon would be difficult to detect. Each region of the galaxy has been enriched by superbubbles $\sim 70$ times. Assuming these enrichment events are independent and identically distributed, then no matter whether the fluctuations are strictly positive, strictly negative, or a combination of the two, the distribution of metallicities around the spatially-varying mean will be approximately Gaussian by the central limit theorem. Therefore, an observed low-metallicity region of gas could be equally-well explained either as the site where a pocket of pristine gas has locally lowered the metallicity; or as a region where fewer superbubble enrichment events have happened to occur. Distinguishing between these two scenarios would require a spatio-temporal statistical analysis that takes into account the metal-enrichment histories of each location within a galaxy. Such data is not available for observed galaxies, but this behaviour could be examined in zoom-in simulations of local galaxies -- for example, in the ISM simulations of \citet{Wibking+Krumholz23}, in which internal metallicity variations by a factor of 0.17 dex in kpc-scale regions \citep{Zhang+25}, roughly in agreement with what is seen in observations. Investigating whether these small-scale fluctuations are seen in numerical models of galaxy evolution and analysing their origins is an intriguing avenue for future research.

It is known that abundances determined using recombination lines do not agree with those made using the direct method \citep{Garcia-Rojas+Esteban07}. 
One possible explanation for this discrepancy is the presence of temperature variations in the interiors of \Hii regions \citep{Peimbert67}.
Such temperature variations may be caused by metallicity variations.
\citet{Stasinska+07} proposed that the temperature fluctuations in the \Hii regions may be caused by small-scale metallicity fluctuations. In their model, a spray of metallic droplets from a galactic fountain fall over the ISM, and these droplets survive until they are ionised within \Hii regions. Based on arguments about their visibility, \citet{Stasinska+07} constrain the radii of these droplets to be smaller than $3 \times 10^{-4}$ pc -- many orders of magnitude smaller than the metallicity fluctuations that are predicted by our model. Therefore, while the metallicity fluctuations that we are proposing can explain the presence of the $\sim100$pc scale, $\sim0.05$ dex metallicity fluctuations that have been observed in recent IFU surveys of nearby galaxies, it does not predict or explain the proposed $\lesssim 1$ pc metallicity fluctuations that may be present in \Hii regions.

With the PHANGS dataset, a moderate disagreement was found between the median size of superbubbles ($\sim 300$ pc), and the scale height of the cold stellar disc of a galaxy ($\sim 200$ pc). This may be because the pressure scale height which sets the size of the superbubbles is slightly different to the scale height of the cold thin stellar disc.
Analysis of more detailed numerical models of superbubble expansion dynamics may be required to answer these questions and determine the exact relationship between a galaxy's scale height and the expected radial extent of superbubbles.

If our model is correct, then the implications for galaxy formation theory are significant, as resolved metallicity observations of galaxies would be able to resolve signatures of face-on superbubbles in a large number of galaxies for which data has already been obtained \citep{Metha+24}. By relating the properties of these superbubbles to a galaxy's global properties, scaling relations could be constructed that allow insights into the connected nature between supernova-driven stellar feedback and the large-scale properties of galaxies. Additionally, examining the sizes of superbubbles in a galaxy would allow the scale height of external galaxies to be measured even when they are completely face-on, providing additional morphological information on the galaxies under investigation. 

Observations with the forthcoming generation of extremely large telescopes will have the capability to capture chemical variations of galaxies on scales finer than 1 pc per pixel \citep{Metha+24}. With such resolution, many more details about the nature of these superbubbles can be revealed. For example, we do not expect all superbubbles to be the same size, as is assumed in our model. Instead, their size is set by the number of supernovae involved in the formation of the superbubble, as well as its relative geometry above or below the galaxy midplane. With sufficient data, a geostatistical analysis may be used to extract the same information that is present in a power spectrum \citep{Rosetta}, which would give the entire spectrum of bubble sizes along with their relative frequencies, providing detailed information about the formation conditions of massive stars.
This data can also be readily converted into a second-order structure function, which provides insights into the nature of turbulence governing chemical mixing in the ISM, as the shape of this function is expected to be different under different turbulent regimes. 
By applying the techniques used in this work to future data sets, more details on the nature of the ISM of galaxies can be uncovered, providing valuable insights into the nature of galaxy evolution.

\section{Summary and conclusions} \label{sec:conclusions}

In this work, we presented a possible explanation for the small ($<0.1$ dex) metallicity fluctuations observed in local face-on spiral galaxies: that the enriched metallicity fluctuations are the face-on signatures of superbubbles that expand outwards until they reach the scale height of the cold component of the ISM. By examining the metallicity of these fluctuations, we find that their progenitor gas clouds had star formation efficiencies of $\epsilon= 0.1-0.2$, indicative of rapid star formation. With this interpretation, we see that star-forming spiral galaxies at $z=0$ have an ISM that is locally bursty. This burstiness is hidden when galaxies are analysed without spatial resolution.

Using the PHANGS-MUSE sample of galaxies, we searched for connections between the local properties of the ISM (the typical spatial extent, $\phi$, and star formation efficiency of superbubbles, $\epsilon$) and a galaxy's global properties. We found that $\phi$ was positively correlated with both the total stellar mass and SFR of a galaxy, whereas $\epsilon$ was positively correlated with only the stellar mass of galaxies. This implies that the local behaviour of the ISM of galaxies is, to some degree, governed by the global properties of a galaxy.

Several predictions are made using our model that will be testable with future studies. By taking a multi-wavelength approach, the presence of these metallicity fluctuations could be correlated against resolved SFR fluctuations, velocity fluctuations, or dust fluctuations seen in infra-red data. Specifically, we predict regions with stronger star formation rates to have higher metallicities, and correlation scales smaller than or equal to the correlation scale for metallicity fluctuations ($\sim 100$ pc). For superbubbles that are still expanding, we expect velocity signatures with sizes of $\sim 100$ pc to be found at regions with local metal enrichment.
We also predict the amount of dust to be lower in regions with large metallicity fluctuations as this dust will have been blown out by superbubbles. The size of the holes in dust regions ought to be similar to the sizes of metal-enriched superbubbles that are found in this work ($\sim 300$ parsecs). Additionally, we expect the size of metallicity fluctuations to increase towards the outer regions of a galaxy's disc as a consequence of disc flaring, and we expect the number density of positive metallicity fluctuations to be higher on the spiral arms of a galaxy compared to in its inter-arm regions. 

If proven to be correct, this theory represents a significant step in our understanding of the interstellar medium of galaxies in a galactic context. Specifically, we have shown that the chemical distributions of galaxies captured by IFU surveys may be used to provide information on the scale heights of galaxies (even when they are face-on) and the star formation efficiencies of \Hii regions. Most importantly, this model presents a compelling scenario of how the metals within galaxies are created and distributed -- from pockets of high mass stars to hot supernovae-driven superbubbles that eventually mix back into the cold ISM. Testing our model predictions will be an exciting avenue of investigation for future works both to demonstrate in action the power of geostatistical analyses and to progress our understanding of the multi-faceted processes that govern galaxy evolution and their chemical enrichment. 







\section*{acknowledgements}

BM acknowledges support from an Australian Government Research Training Program (RTP) Scholarship and The David Lachlan Hay Memorial Fund. BM would further like to thank Dr. Katie Auchettl for helpful conversations about core collapse supernovae. This research is supported in part by the Australian Research Council Centre of Excellence for All Sky Astrophysics in 3 Dimensions (ASTRO 3D), through project number CE170100013. The majority of this research was conducted on Wurundjeri land.

\section*{data availability}
All derived data products used in this analysis are available from the corresponding author upon reasonable request. Original emission line maps from the PHANGS survey are available from the PHANGS website: \url{http://www.phangs.org/data}.

\bibliographystyle{mnras}
\bibliography{biblio} 

\newpage
\appendix

\section{Analysis with alternative metallicity diagnostics}
\label{ap:other_Z_diags}

To ensure our analysis is robust to the choice of metallicity diagnostic, we repeat our analysis with two alternative metallicity diagnostics: \NSH, calibrated using the \Hii region modelling code \textsc{Mappings-5.0} \citep{Dopita+16}, and the S-calibration (or Scal for short) of \citet{Pilyugin+Grebel16}, calibrated emprically using $T_e$-based metallicity estimates of 313 nearby \Hii regions.
We note that the \NSH\ diagnostic is largely insensitive to ionisation parameter \citet{Kewley+19}.

\begin{figure*}
    \centering
    \includegraphics[width=\linewidth]{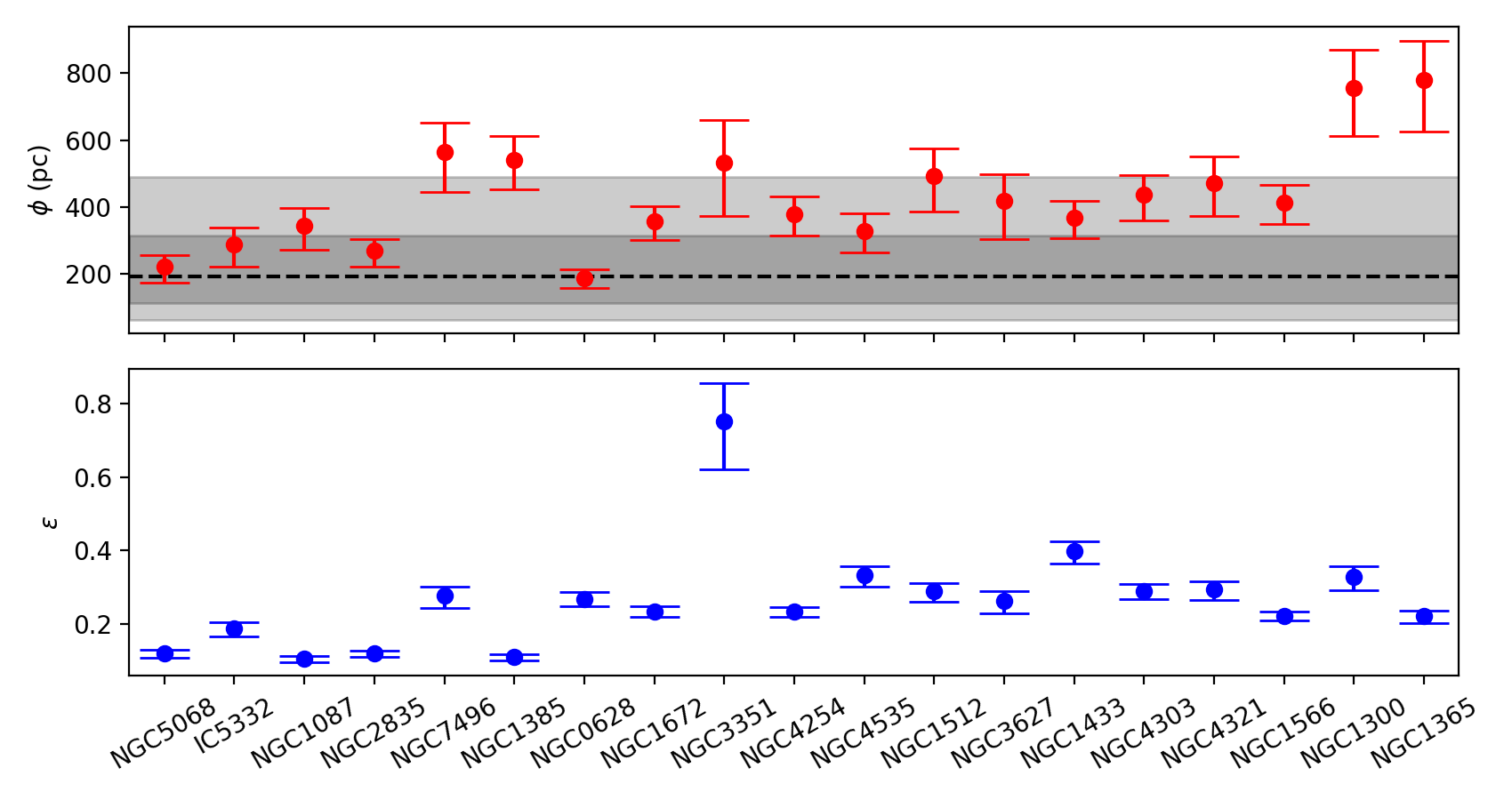}
    \caption{As in Figure \ref{fig:smallscale_params}, but with \NSH\ as the metallicity diagnostic.}
    \label{fig:results_NSH}
\end{figure*}

We show the recovered properties of both $\epsilon$ and $\phi$ for all galaxies when \NSH\ is used as the metallicity diagnostic in Figure \ref{fig:results_NSH}. When \NSH\ is used, we find that values of $\phi$ are typically larger than the typical scale heights of the thin disc, with a median value of $412$ pc across the population. In agreement with the results when \ON\ is used, the two largest galaxies by mass (NGC 1300 and NGC 1365) have the largest values of $\phi$, followed by NGC 7496 and NGC 1385. With this diagnostic, $95\%$ of the sample display $\epsilon$ values of $0.1-0.4$, with one outlier (NGC 3351) displaying an extreme $\epsilon$ of $0.75^{+0.13}_{-0.11}$. Even accounting for this outlier, the median value of $\epsilon$ over the population when \NSH\ was used was found to be $0.26$, consistent with enhanced star formation occurring at the sites of positive metallicity fluctuations.

\begin{figure*}
    \centering
    \includegraphics[width=\linewidth]{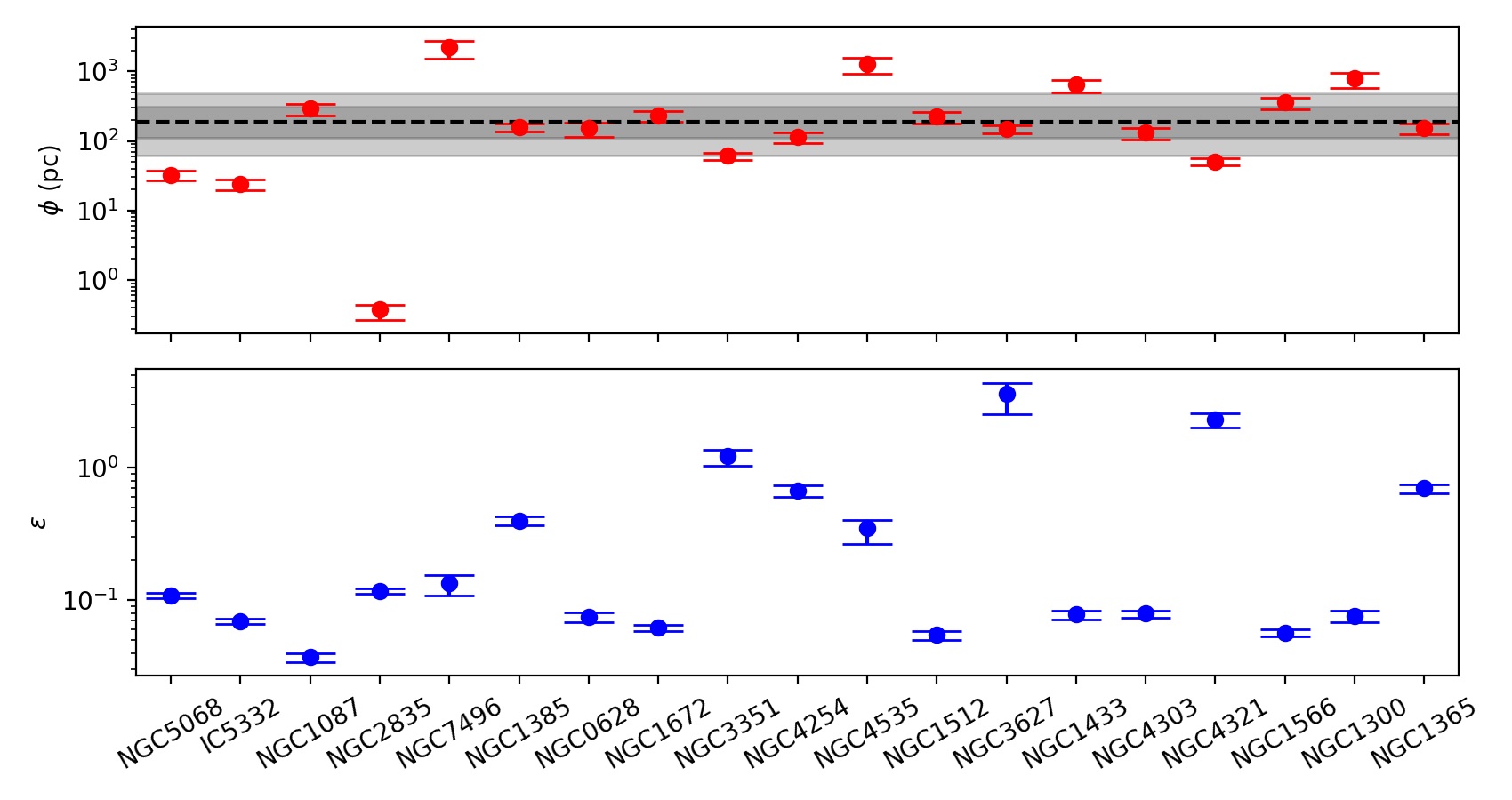}
    \caption{As in Figure \ref{fig:smallscale_params}, but with Scal as the metallicity diagnostic. Note that this plot uses log-scaling on the $y$-axes due to the large range of values of $\epsilon$ and $\phi$ found when this diagnostics was used.}
    \label{fig:results_Scal}
\end{figure*}

When the Scal diagnostic was used, several numerical issues occurred. We plot the revocered values of $\phi$ and $\epsilon$ for all galaxies when Scal is used in Figure \ref{fig:results_Scal}. Both extremely high ($\phi = 2202^{+701}_{-565}$ pc for NGC 7496) and extremely low ($\phi = 0.38^{+0.11}_{-0.05}$ pc for NGC 2835) values of $\phi$, outside of the soft bounds of our prior, were recovered for galaxies when this diagnostic was used. Despite the wide range in recovered values, the median value of $\phi$ across the sample when Scal was used was found to be $154$ pc, comparable to the scale heights of cold stellar discs. Turning to $\epsilon$, we also found a wide range of values, from $\epsilon= 0.037^{+0.003}_{-0.002}$ for NGC 1087, to three unphysical values of $\epsilon > 1$ for NGC 3351, NGC 3627, and NGC 4321. Again, despite the wide range of recovered values of $\epsilon$, we find the median value across the analysed population to be $\epsilon= 0.11$, in line with our model predictions. Therefore, this data, at least on average, agrees with our model.

We investigated several possible reasons for the observed wide spread in recovered ISM parameters when Scal was used.
The values of the S/N ratio for metallicities computed with Scal were not found to be higher than those measured with \NSH. Similarly, the overall metallicities of galaxies when Scal was used was found to be lower than those calculated with \ON\ or \NSH, which should act to lower the values of $\epsilon$ recovered rather than generate unphysically high values. Every emission line used to compute Scal is also used to either compute \ON\ or \NSH, meaning that an underestimation on the errors of any of the emission lines ought to affect multiple diagnostics and not just Scal. We cannot rule out the fact that such variance in recovered parameters is caused by model error -- that is, that the Scal metallicity diagnostic is less accurate than the \ON\ or \NSH\ diagnostics on finely spatially-resolved data.




\label{lastpage}
\end{document}